\documentclass[%
 aip,
 amsmath,amssymb,
reprint, onecolumn 
]{revtex4-1}

\usepackage{graphicx}
\usepackage{dcolumn}
\usepackage{bm}

\usepackage[utf8]{inputenc}
\usepackage[T1]{fontenc}
\usepackage{mathptmx}
\usepackage{etoolbox}

\newcommand{\bra}{\langle}
\newcommand{\ket}{\rangle}

\newcommand{\dd}[2]{\frac {\partial #1}{\partial #2}}

\newcommand{\DD}[2]{\frac {d #1}{d #2}}

\newcommand{\beq}{\begin{equation}}
\newcommand{\eeq}{\end{equation}}
\newcommand{\beqs}{\begin{eqnarray}}
\newcommand{\eeqs}{\end{eqnarray}}

\newcommand{\ov}[1]{\frac{1}{#1}}
\newcommand{\fr}[2]{\frac{#1}{#2}}

\def\al{\alpha}		 		 
\def\del{\delta}			 
			 
\def\la{\lambda}		\def\sig{\sigma}		
		\def\tht{\theta}	
		\def\om{\omega}		\def\Om{\Omega}

\usepackage{amsmath,amssymb} 

\usepackage{calligra} 
\DeclareMathAlphabet{\mathcalligra}{T1}{calligra}{m}{n}
\DeclareFontShape{T1}{calligra}{m}{n}{<->s*[2.2]callig15}{}

\newcount\colveccount  
\newcommand*\colvec[1]{\global\colveccount#1  \begin{pmatrix} \colvecnext} \def\colvecnext#1{#1 \global\advance\colveccount-1
        \ifnum\colveccount>0 \\ \expandafter\colvecnext
        \else \end{pmatrix} \fi}

\usepackage[skip=3pt,labelfont=normal,labelsep=colon]{caption}
\usepackage[skip=3pt]{subcaption} 
\usepackage{wrapfig}

\usepackage{bm}

\makeatletter
\def\@email#1#2{%
 \endgroup
 \patchcmd{\titleblock@produce}
  {\frontmatter@RRAPformat}
  {\frontmatter@RRAPformat{\produce@RRAP{*#1\href{mailto:#2}{#2}}}\frontmatter@RRAPformat}
  {}{}
}%
\makeatother

\begin{document}


\title[Poisson Geometric Formulation of Quantum Mechanics]{
Poisson Geometric Formulation of Quantum Mechanics}
\author{Pritish Sinha}
\author{Ankit Yadav}
 \email{pritish@cmi.ac.in}
\affiliation{Chennai Mathematical Institute,  SIPCOT IT Park, Siruseri 603103, India}
 \date{4 March 2024}

\begin{abstract}
    We study the Poisson geometrical formulation of quantum mechanics for finite dimensional mixed and pure states. Equivalently, we show that quantum mechanics can be understood in the language of {\it classical mechanics}. We review the symplectic structure of the Hilbert space and identify its canonical coordinates. We extend the geometric picture to the space of density matrices $D_N^+$. We find it is not symplectic but admits a linear $\mathfrak{su}(N)$ Poisson structure. We identify Casimir surfaces of $D_N^+$ and show that the space of pure states $P_N \equiv \mathbb{C}P^{N-1}$ is one of its symplectic submanifolds which is an intersection of primitive Casimirs. We identify generic symplectic submanifolds of $D_N^+$ and calculate their dimensions. We find that $D_N^+$ is singularly foliated by the symplectic leaves of varying dimensions, also known as coadjoint orbits. We also find an ascending chain of Poisson submanifolds $D_N^M \subset D_N^{M+1}$ for $ 1 \leq M \leq N-1$. Each such Poisson submanifold $D_N^M$ is obtained by tracing out the $\mathbb{C}^M$ states from the bipartite system $\mathbb{C}^N \times \mathbb{C}^M$ and is an intersection of $N-M$ primitive Casimirs of $D_N^+$. Their Poisson structure is induced from the symplectic structure of the bipartite system. We also show their foliations. Finally, we study the positive semi-definite geometry of the symplectic submanifold $E_N^M$ consisting of the mixed states with maximum entropy in $D_N^M$.

\end{abstract}  

\maketitle

\normalsize

\section{Introduction}
\label{s:intro}

Classical mechanics is best understood in its geometrical picture: symplectic formulation of classical mechanics where we have a phase space of states endowed with a symplectic 2-form and a Hamiltonian function generating the time evolution. On the other hand, quantum mechanics is usually studied in algebraic or the functional-analytic formulation. However, it turns out that there is an equivalent geometrical description of quantum mechanics which also admits the symplectic structure. In Ref.~[\onlinecite{Kibble}], it was shown that the Projective Hilbert space of pure states is a symplectic manifold and the Schr\"odinger equation can be interpreted as Hamilton's equation where Hamiltonian function is simply the expectation of Hamiltonian operator. In Ref.~[\onlinecite{Hilso}], it was found for finite dimensional quantum mechanics that it admits a K\"ahler structure, naturally endowed by the Hermitian inner-product of the Hilbert space. This allowed quantum mechanics to be interpreted as a generalized classical mechanics with analogous symplectic picture. However, unlike classical mechanics, quantum mechanical phase space also admits a compatible Riemannian structure, providing a natural notion of distance between the quantum mechanical states and interpretation of Planck's constant in terms of curvature of the projective Hilbert space. 

 A detailed geometric formulation of the postulates of quantum mechanics was given in Ref.~[\onlinecite{AbAs-Tras}] for infinite dimensional Hilbert space. Their geometric picture presents a sharp contrast between the structural similarities and differences of classical and quantum mechanics. The symplectic structure is responsible for the features of quantum mechanics that have direct classical analogues such as observable being represented by real-valued function on the quantum phase space and the Schrödinger evolution being captured by the symplectic flow generated by a Hamiltonian function. On the other hand, differences between classical and quantum mechanics are geometrically reflected by the existence of Riemannian structure in quantum phase space which is absent in the classical phase space. In classical mechanics, measurement of an observable is simply the value of the function at the point and the state is not disturbed by it. There is also no uncertainty in the occurrence of outcomes. Whereas the quantum mechanical processes further involve probabilistic interpretation, quantum uncertainties and state vector reduction in a measurement process. And it is shown that they can all be formulated in terms of compatible Riemannian metric. The authors also used this geometric picture to classify possible generalizations of quantum mechanics as well as to address issues in second quantization and semi-classical approximations \cite{AbAs-Tras, geo-qua-mek}. 
 
 The geometric formulation enables to construct the quantum phase space as a fibre bundle over the base space: classical phase space \cite{AbAs-Tras, Bojowald-Skirzewski-eff}. In Ref.~[\onlinecite{Bojowald-Skirzewski-eff}], expectations and Weyl-ordered moments of quantum states are used to furnish coordinates for fibre bundle phase space. Using the symplectic structure of the phase space, commutation relations between the coordinates are obtained. This makes it possible to express Schrödinger equation as classical-like Hamilton's equations involving infinitely many variables. It reduces to exact classical Hamilton's equation when quantum fluctuations are ignored. When truncated at some non-zero finite order of fluctuation, Schrödinger equation reduces to finite dimensional effective equations of motion describing appropriate semi-classical regime. They become useful in many different areas such as quantum chemistry\cite{Prez-QC} and quantum cosmology\cite{Boj-QC}. The method extends the classical phase space to a higher dimensional (non-symplectic) Poisson manifold in which the expectation values are coupled to moments of up to some finite order. A systematic method for semi-classical truncation to obtain effective equations of motion and the Poisson geometry of the semi-classical phase space are studied in Ref.~[\onlinecite{Bojowald-Baytas-Crowe}]. In particular, faithful canonical realizations of the Poisson algebra are constructed for certain semi-classical phase spaces. For $2N$ classical degrees of freedom, it is found that semi-classical truncation to the second order moments constitutes an $N(2N+3)$ dimensional semi-classical phase space which has $2N$ usual classical degrees of freedom and $N(2N+1)$ second order moments that form $\mathfrak{sp}(2N)$ Lie algebra. The method for obtaining their Casimir–Darboux coordinates is also discussed in Ref.~[\onlinecite{Bojowald-Baytas-Crowe}].
 
 The geometrical formulation of quantum mechanics developed above does not include the mixed states. As suggested in Ref.~[\onlinecite{Bojowald-Baytas-Crowe}], the effective methods can be extended to include mixed states as well. In this paper, we show the symplectic geometric formulation of quantum mechanics for pure states naturally extends to a Poisson geometric (non-symplectic) formulation of quantum mechanics for density matrices in finite dimensions. Hereby, we present a brief outline of our paper. 

In Section \ref{s:symplectic-pure-states}, we review the symplectic geometric formulation of quantum mechanics for finite dimensional space of pure-states. In Section \ref{s:mixed-states}, we show that the space of density matrices $D_N^+$ admits a linear Poisson structure (also known as Kirillov-Kostant-Souriau structure) inherited by the $\mathfrak{su}(N)$ lie algebra of Hermitian matrices. The Hamiltonian function generating the dynamics of density matrices is the expectation of Hamiltonian operator Tr$(\rho 
\hat{H})/2 \hbar$ defined on the full space of density matrices. In Section \ref{s:casimir-invariants}, we show that $D_N^+$ has degenerate Poisson structure and we classify its Casimir surfaces. The space of pure states $P_N \equiv \mathbb{C}P^{N-1}$ is shown as an intersection of primitive (algebraically independent) Casimirs. Poisson structure on the space of density matrices $D_N^+$ is reconciled with the symplectic structure on the space of pure states $P_N$. The positive semi-definite constraint on the Casimir surfaces is discussed as this is what distinguishes physical density matrices from general Hermitian matrices of unit trace. We show that the space of density matrices is singularly foliated by its symplectic leaves. The leaves are also known as coadjoint orbits generated due to conjugate actions by Unitary operators on density matrices. The dimensions of the leaves of $D_N^+$ are calculated. In Section \ref{s:trace-geo}, we study the Poisson geometry of the reduced density matrix space $D_N^M$ arising from symplectic structure in $\mathbb{C}^N \times \mathbb{C}^M$ via partial tracing of its (pure) states. We find an ascending chain of Poisson submanifolds $D_N^M \subset D_N^{M+1}$ with respect to the inclusion $\subset$ for $1 \leq M \leq N-1$, each $D_N^M$ being an intersection of $N-M$ Casimirs in $D_N^+$. Their foliations are also shown. As $M$ is increased, $D_N^M$ admits symplectic leaves containing mixed states with larger and larger entropy. Finally, we study the positive semi-definite geometry of the symplectic submanifold consisting of mixed states with largest entropy in $D_N^M$.   

\section{Geometry of Pure states}
\label{s:symplectic-pure-states}

We start with the finite dimensional Hilbert space ${\cal H}_{N} =  \mathbb{C}^N$. Dynamical system with only spin  degrees of freedom are described by states which belong to ${\cal H}_{N}$. $\mathbb{C}^N$ is also a symplectic manifold with K\"ahler symplectic structure. 

Motivated by the connection between infinite dimensional Hilbert space in quantum mechanics and the associated phase space parameterized by expectation value of Weyl-ordered operators \cite{Bojowald-Skirzewski-eff}, in this section we define canonical coordinates for ${\cal H}_{N}$. This does not allow us to use expectation of Hermitian operators to construct the canonical coordinates. To show this, let's assume finite dimensional phase space admits canonical variables in the form of expectations of Hermitian operators. Then there exist $N \times N$ Hermitian operators $A$ and $B$ such that
\beq
\{\bra A \ket, \bra B \ket \} = \bra [A,B] \ket = 1 \:\:\forall \:\: \text{points in phase space.}
\eeq
Now, this  is only possible if $[A,B] = I$. Since A and B are Hermitian, they may be written as sum of generalized Gell-Mann matrices. Thus, we may write $A =  a_0 I + a_k T_k$ and $B = b_0 I + b_k T_k$, where $a_i$, $b_i$s are real and $T_k$s are generalized Gell-Mann matrices. Gell-Mann matrices satisfy the following commutation relation
\beq
[T_a, T_b] = i f_{abc}T_c.
\eeq
This implies
 \beq
[A,B] = a_i b_j [T_i, T_j] = a_i b_j f_{ijk} T_k.
\eeq

Taking the trace, we get $\text{Tr}([A,B]) = a_i b_j f_{ijk} \text{Tr}(T_k) = 0$. Thus, their commutation relation cannot be a multiple of identity. Hence, Canonical variables on a finite dimensional phase space cannot be expressed as expectations of Hermitian operators. 

There is yet another way to obtain canonical variables which will be convenient for what follows.

Consider 2 vectors $a = (a_1, a_2, \ldots, a_N)$ and $b = (b_1, b_2, \ldots, b_N)$ in $\mathbb{C}^N$. The inner-product in $\mathbb{C}^N$ is defined as follows:

\beq
\bra{b|a}\ket = \sum_i b_i^* a_i = \sum_i (a_i^R b_i^R+a_i^I b_i^I) + i (a_i^I b_i^R-a_i^R b_i^I ).
\eeq
 We can separate real and imaginary parts of the inner product to define
\beq
G(a,b) =  \text{Re} \bra{b|a}\ket = (a_1^R, \ldots, a_N^R, a_1^I, \ldots, a_N^I)  
\begin{pmatrix}
I & 0 \\
0 & I 
\end{pmatrix}
(b_1^R, \ldots, b_N^R, b_1^I, \ldots,b_N^I)^T
\eeq
 
and 

\beq
\Om(a,b) =\text{Im} \bra{b|a}\ket = (a_1^R, \ldots, a_N^R, a_1^I, \ldots, a_N^I) 
\begin{pmatrix}
0 & -I \\
I & 0 
\end{pmatrix}
(b_1^R, \ldots, b_N^R, b_1^I, \ldots, b_N^I)^T
\label{eq:symplectic-product}
\eeq

We see $G(a,b)$ and $\Om(a,b)$ form a metric and symplectic structure respectively on $\mathbb{R}^{2N}$, induced from the inner product structure on the complex vector space. 
We also note from the definition, metric and symplectic structures are related by $G(a,b) = \Om(a,J b)$, where $J = \begin{pmatrix} 0 & -I \\ I & 0  \end{pmatrix}$ is the complex structure on $\mathbb{R}^{2N}$. $(J,G,\Om)$ equips $\mathbb{R}^{2N}$ with a K\"ahler structure.

From (\ref{eq:symplectic-product}), we observe that the real and imaginary parts of complex coordinates naturally provide us with canonically conjugate pairs on $\mathbb{R}^{2N}$. Since the symplectic 2-form $\om_{ij} = J = \begin{pmatrix} 0 & -I \\ I & 0  \end{pmatrix}$ and $J^2 = -I$, we can define the poisson tensor $r^{ij} = -J = \begin{pmatrix} 0 & I \\ -I & 0  \end{pmatrix}$. Consider a complex vector $z = (x_1 + i y_1, \cdots, x_N + i y_N)$, then the canonical coordinates on $\mathbb{R}^{2N}$ are $(x_1, \cdots, x_N, y_1, \cdots, y_N)$ with $\{x_m,y_n\} = -\{y_n,x_m\} = \del_{mn}$ whereas $\{x_m,x_n\} = \{y_m,y_n\}  = 0$.

 In  geometric formulation of quantum mechanics $\bra \hat H \ket/2  \hbar$ is the Hamiltonian function, where $\hat H$ is some self-adjoint Hamiltonian operator. This is simply because, the Hamilton's equation for this Hamiltonian wrt (\ref{eq:symplectic-product}) is the Schr\"odinger's equation as we now prove : 
\beq
i  \hbar \dd{z_a}{t} = \hat H_{ab}z_b.
\eeq
We write Schr\"odinger's equation in canonical coordinates  as follows:
\beq
\begin{pmatrix}
 \dot{x_a} \\
 \dot{y_a} 
\end{pmatrix} =
\frac{1}{\hbar}\begin{pmatrix}
\hat H_{ab}^I & \hat H_{ab}^R \\
 - \hat H_{ab}^R & \hat H_{ab}^I 
\end{pmatrix}  
\begin{pmatrix}
 x_b \\
 y_b 
\end{pmatrix}.
\label{eq:canoschrod}
\eeq
The canonical equations of motion, on the other hand can be written as , 
\beq
\dot{\xi^i }=\{\xi^i, H\}= r^{ij} \dd{H}{\xi^j}.
\label{eq:Hamiltoneq}
\eeq
where $\xi^i$ are canonical coordinates and $H = \frac{\bra \hat H \ket} { 2 \hbar} =\frac{ \hat H_{ab}}{2 \hbar} z_a^* z_b $:
\beq
H = \frac{1}{2 \hbar}[\hat H^R_{ab}(x_a x_b + y_a y_b) - \hat H^I_{ab}(x_a y_b - y_a x_b)] + i \underbrace{[\hat H^I_{ab}(x_a x_b + y_a y_b) + \hat H^R_{ab}(x_a y_b - y_a x_b)] .}_{=0} 
\eeq

The imaginary part is 0 as expected from a Hermitian operator ($\hat H^R_{ab} = \hat H^R_{ba}$ and $\hat H^I_{ab} = - \hat H^I_{ba}$).  Taking partial derivative wrt canonical variables and using Hermiticity, we get
\beq
\dd{H}{\xi^j} = \begin{pmatrix}
 \dd{H}{x_a} \\
 \dd{H}{y_a} 
\end{pmatrix} =
\frac{1}{\hbar}\begin{pmatrix}
 \hat H_{ab}^R & -\hat H_{ab}^I \\
 \hat H_{ab}^I & \hat H_{ab}^R 
\end{pmatrix}  
\begin{pmatrix}
 x_b \\
 y_b 
\end{pmatrix}.
\eeq
Rewriting (\ref{eq:canoschrod}) as follows 
\beq
\begin{pmatrix}
 \dot{x_a} \\
 \dot{y_a} 
\end{pmatrix} =
\frac{1}{\hbar}\begin{pmatrix}
 \hat H_{ab}^I & \hat H_{ab}^R \\
 -\hat H_{ab}^R & \hat H_{ab}^I 
\end{pmatrix}  
\begin{pmatrix}
 x_b \\
 y_b 
\end{pmatrix}
=
\begin{pmatrix} 0 & I \\ -I & 0  \end{pmatrix}
\frac{1}{\hbar}\begin{pmatrix}
 \hat H_{ab}^R & -\hat H_{ab}^I \\
 \hat H_{ab}^I & \hat H_{ab}^R 
\end{pmatrix}  
\begin{pmatrix}
 x_b \\
 y_b 
\end{pmatrix}.
\eeq
 which is exactly the Hamilton's equation (\ref{eq:Hamiltoneq}). 

Thus, the so-called geometric formulation of quantum mechanics can be applied even in the case of  finite dimensional quantum mechanical systems: The resulting symplectic form can be recasted in Darboux form where the canonically conjugate pairs are $\{ x_{a},\, y_{a}\vert\, a = (1, \dots, N)\}$. It is important to emphasize that the Hilbert space considered above is $\mathbb{C}^N$, whereas the physical Hilbert space is the projective Hilbert space $\mathbb{C}P^{N-1}$.

\section{Geometry of Mixed states}
\label{s:mixed-states}

In this section, we extend the geometry to include finite dimensional mixed states as well. As is well known, all the states (pure as well as mixed) can be represented by density matrices $\rho$ which is a positive semi-definite Hermitian operator of unit trace acting on the Hilbert space of the system. An $N \times N$ density matrix can be written as

\beq
\rho = \frac{I}{N} + \sum_{k=1}^{N^2-1} \al_k T_k
\label{eq:dens-defn}
\eeq
where $\al_k$ are real and $T_k$ are complex $N \times N$ Gell-Mann matrices. $T_{k}$s are Hermitian and trace-free.
\beq
T_aT_b = \frac{1}{2N} \del_{ab} I_N + \frac{1}{2} \sum_{c=1}^{N^2-1} (i f_{abc}+d_{abc})T_c
\label{eq:prodTT}
\eeq
where $f_{abc}$ are the structure constants, while $d_{abc}$ are known as symmetric coefficients. Since any $N\times N$ density matrix $\rho$ can be represented by (\ref{eq:dens-defn}), we can assign $(\al_1,\al_2, \cdots,\al_{N^2-1})$ as coordinates to the space of all possible $N\times N$ density matrices. We call these as Gell-Mann coordinates. Note the (\ref{eq:dens-defn}) need not necessarily represent a density matrix for an arbitrarily chosen real values of $\al_i$s. Rather they represent all Hermitian matrices with trace $1$. For it to qualify as a density matrix, (\ref{eq:dens-defn}) must also be positive semi-definite. 

The set of all $N \times N$ Hermitian matrices of unit trace constitute a real-manifold of $N^2-1$ dimensions. We call this space $D_N$. We denote $D_N^+ \subset D_N$ as a $N^2-1$ dimensional subspace of positive semi-definite Hermitian matrices of unit trace qualifying it to be the space of physical density matrices. We will return to how positive semi-definiteness constrains the geometry of the state-space. For the moment, we focus on finding a possible Poisson geometric structure on $D_N^+$. The structure does not assume any positive semi-definiteness, so we do not need to make any explicit distinction between  $D_N^+$ and $D_N$. When required, we will bring up the positive semi-definiteness.
 We start with a Hamiltonian operator $\hat{H}$, with respect to which density matrices evolve as
\beq
\DD{\rho}{t} = \fr{i}{\hbar} [\rho,\hat{H}].
\eeq
Since $\hat{H}$ is also self-adjoint operator, we can write 
\beq
\hat{H} = h_0 I +\sum_{k=1}^{N^2-1} h_k T_k.
\eeq
Thus,
\beq
\dot{\rho} = \dot{\al_i} T_i =  \fr{i}{\hbar} [\rho,\hat{H}] = \frac{i}{\hbar} \al_i h_j [ T_i, T_j ] = -\frac{1}{\hbar} f_{ijk} \al_i h_j T_k. 
\eeq
Multiplying by $T_l$ and taking trace both sides, we get
\beq
\dot{\al_l} = -\fr{1}{\hbar} f_{lij} \al_i h_j. 
\label{eq:mixed-unitary}
\eeq
For relation (\ref{eq:mixed-unitary}) to qualify as a Hamiltonian time evolution on the phase space $D_N$, there has to exist a Poisson structure $ \left( \{,\}:C^{\infty}(D_N)\times C^{\infty}(D_N) \to C^{\infty}(D_N)\right) $ and a function $H(\al)$ on the space $D_N$ such that $\dot{\al} = \{\al, H\}$. We claim that $D_N$ is equipped with a Poisson structure
\beq
 \{\al_i,\al_j\} = r^{ij}
\label{eq:Mixed-Poisson-Tensor}
\eeq
and the Hamiltonian function on $D_N$ is
\beq
H(\al) = \fr{1}{2 \hbar}\text{Tr}(\rho \hat{H}) =\fr{1}{2 \hbar} \sum_{i,j =0}^{i,j = N^2-1}\al_i h_j \text{Tr}( T_i T_j) =\fr{1}{2 \hbar}\left[ h_0 +\sum_{k =1}^{k = N^2-1} \frac{\al_k h_k}{2} \right].
\label{eq:Ham-DN}
\eeq
Using (\ref{eq:Mixed-Poisson-Tensor}) and (\ref{eq:Ham-DN}), we get the Hamiltonian equations of motion
\beq
\dot{\al_l} =  \ov{2\hbar}\{\al_l, \text{Tr}(\rho \hat{H}) \} = r^{lj} \frac{h_j}{4 \hbar}.
\label{eq:Mixed-Poisson}
\eeq
Comparing (\ref{eq:mixed-unitary}) and (\ref{eq:Mixed-Poisson}), we get
\beq
 r^{ij} = \{\al_i,\al_j\} = 4 f_{ijk} \al_k,
\label{eq:suN-Poisson-Tensor}
\eeq
which satisfies all properties of Poisson tensor. The second equality in fact shows that $\{ \alpha_{l}\}$ generate an $\mathfrak{su}(N)$ Lie algebra. The tensor $r$ is antisymmetric and  satisfies Lebiniz rule and Jacobi relation:
 \beq
 r^{lk} \partial_l r^{ij} + r^{li} \partial_l r^{jk} + r^{lj} \partial_l r^{ki} = 16(f_{lkm}f_{lij}+f_{lim}f_{ljk}+f_{ljm}f_{lki})\al_m = 0 
\label{eq:Jacobi}
\eeq  
 Hence, $D_N$ is an $N^2-1$ dimensional real manifold equipped with a Poisson tensor $r^{ij}$ (\ref{eq:suN-Poisson-Tensor}). 

\section{Casimir Invariants}
\label{s:casimir-invariants}
 
In this section, we explore implications of the Poisson structure on $D_N$. $D_N$ is a Poisson but not symplectic as, $r$ in (\ref{eq:suN-Poisson-Tensor}) is degenerate.  We classify here the level surfaces of the Casimirs whose intersections are symplectic submanifolds. Interestingly enough, pure states constitute one of the symplectic submanifolds of $D_N$. If $F$ is a Casimir, then it commutes with all $\al_i$s such that

\beq
\{\al_i,F\} = r^{ij} \dd{F}{\al_j} = 4 f_{ijk} \al_k \dd{F}{\al_j}  = 0.
\label{eq:Casimir}
\eeq 
 
If $F = \al_l\al_l$, the Poisson commutation relation $\{\al_i,F\} = r^{ij} \al_j = 4 f_{ijk} \al_k \al_j = 0$ since $f$ is antisymmetric in all indices. Thus $F = \sum_{l=1}^{l=N^2-1}\al_l\al_l$ is a quadratic Casimir. This also establishes the degeneracy of $r^{ij}$. For $N>2$, $D_N$ also admits higher order Casimirs. 
A cubic Casimir, for example, is $F=d_{ijk} \al_i \al_j \al_k$. The commutation relation for cubic Casimir is
\beq
\{\al_l, F\} = 4 f_{ljk} \al_k (d_{jnp} \al_n \al_p + d_{mjp} \al_m \al_p + d_{mnj} \al_m \al_n). 
\label{eq:cubic}
\eeq 

Changing $k \leftrightarrow m$, $k \leftrightarrow n$ and $k \leftrightarrow p$ and doing cycling permutation for indices, we get 
\beq
\{\al_l, F\} = 4 ( f_{mlj} d_{npj}  + f_{nlj} d_{pmj}  +  f_{plj} d_{mnj})  \al_m \al_n \al_p = 0.
\label{eq:}
\eeq
The above follows from second Jacobi identity \cite{kaplan} (see Appendix \ref{s:ad-inv}).

For any finite-dimensional semisimple Lie algebra, the number of primitive (i.e. algebrically independent) Casimirs is equal to the rank of the semisimple Lie algebra\cite{Xav}. Hence, $D_N$, whose coordinates generate $\mathfrak{su}(N)$ Lie algebra, has $N-1$ primitive casimirs. An $m$-order Casimir for $\mathfrak{su}(N)$ can be written in terms of symmetric coefficients \cite{a-j-m-inv-ten,Az-Ma-Mo-Bu} as
\beq
C^{(m)} = g_{i_1 i_2 \cdots i_m} \al_{i_1} \al_{i_2} \ldots \al_{i_m}
= {d_{(i_1i_2}}^{j_1} {{d^{j_1}}_{i_3}}^{j_2}{{d^{j_2}}_{i_4}}^{j_3}\ldots{{d^{j_{m-4}}}_{i_{m-2}}}^{j_{m-3}}{d_{i_{m-1}i_m)}}^{j_{m-3}} \al_{i_1} \al_{i_2} \ldots \al_{i_m}.
\label{eq:m-order-casimir}
\eeq
The $()$ denotes a symmetrized sum in $i_1, i_2, \cdots, i_m$ indices.  The symmetric tensors $g_{i_1 i_2 \cdots i_m}$ (\ref{eq:m-order-casimir}) constitute what is known as Sudbery basis \cite{sudbery} for symmetric invariant tensors of $\mathfrak{su}(N)$.
 They satisfy the ad-invariance identity (for proof see Appendix \ref{s:ad-inv})
\beq
\sum_{s=1}^{m} f_{\nu i_s \rho} g_{i_1 \ldots \hat{i}_s\: \rho \: i_{s+1} \ldots i_m} = 0,
\label{eq:G-invariance-identity}
\eeq
which precisely makes its commutation with any $\al_i$ as 0, qualifying $C^{(m)}$ as a Casimir. If $m > N$, the higher order Casimirs can be decomposed in terms of primitive Casimirs of order $m \leq N$ \cite{a-j-m-inv-ten}. 
We will refer to the collection of Casimirs $\{C^{(m)}\}$ as Sudbery basis Casimirs.

\subsection{Pure state Casimir surfaces in Sudbery basis}
\label{Pure-stat-casimir}
We characterize the pure state space ($P_N$) as a subspace of $D_N$. Pure states satisfy $\rho^2 = \rho$. 
\beq
\rho^2 =  I\left(\frac{1}{N^2}+\frac{\al_k\al_k}{2N}\right) + T_k \left( \frac{2 \al_k}{N} + \frac{\al_i\al_j}{2} d_{ijk} \right).
\label{eq:rho-square}
\eeq
Equating this to (\ref{eq:dens-defn}), we get the equation for surface:
\beq
\al_k \al_k = 2\bigg(1-\frac{1}{N}\bigg)
\label{eq:pure-state-surface}
\eeq
and 
\beq
\al_k \bigg(1-\frac{2}{N}\bigg) = \frac{\al_i \al_j}{2} d_{ijk}.
\label{eq:pure-state-surface-2}
\eeq
Contracting (\ref{eq:pure-state-surface-2}) with $\al_k$ and using (\ref{eq:pure-state-surface}), we get
\beq
4\bigg(1-\frac{1}{N}\bigg) \bigg(1-\frac{2}{N}\bigg) =  d_{ijk} \al_i \al_j\al_k .
\label{eq:pure-state-surface-3}
\eeq
It implies that the pure states always lie on the intersection of second order and third order Casimir surfaces given by equation (\ref{eq:pure-state-surface}) and (\ref{eq:pure-state-surface-3}). In fact, any pure state necessarily lies on the intersection of all primitive Casimir surfaces. We take equation (\ref{eq:pure-state-surface}) and contract it with products of $d_{ijk}$ and the phase space variable $\al_i$s on both sides and relabel to get
\beq
\text{R.H.S.}\:   = \:  \frac{1}{2} {d_{i_1i_2}}^{j_1} {{d^{j_1}}_{i_3}}^{j_2}{{d^{j_2}}_{i_4}}^{j_3} \ldots {{d^{j_{m-4}}}_{i_{m-2}}}^{j_{m-3}}\underbrace{{d_{i_{m-1}i_m}}^{j_{m-3}} \al_{i_{m-1}}\al_{i_m}}_{d_{ijk}\al_i\al_j \:\:\text{in (\ref{eq:pure-state-surface-2})}}\al_{i_1} \al_{i_2} \ldots \al_{i_{m-2}}
\label{eq:RHS-casimir-dijk}
\eeq
and
\beq
\text{L.H.S.}\:   = \:
\bigg(1-\frac{2}{N}\bigg) {d_{i_1i_2}}^{j_1} {{d^{j_1}}_{i_3}}^{j_2}{{d^{j_2}}_{i_4}}^{j_3} \ldots {{d^{j_{m-5}}}_{i_{m-3}}}^{j_{m-4}} {{d^{j_{m-4}}}_{i_{m-2} i_{m-1}}}
 \underbrace{\al_{i_{m-1}}}_{\al_k \:\:\text{in (\ref{eq:pure-state-surface-2})}}\al_{i_1} \al_{i_2}.
\ldots \al_{i_{m-2}}.
\label{eq:LHS-casimir-dijk}
\eeq
Since variables $\al_i$s multiply commutatively, any permutation of $(i_1, i_2, i_3, \ldots)$ will not change the equations (\ref{eq:RHS-casimir-dijk}) and (\ref{eq:LHS-casimir-dijk}). This allows us to write the above equation as symmetrized sum:
\beqs
   \frac{1}{ (m-1)!} \bigg(1-\frac{2}{N}\bigg) {d_{(i_1i_2}}^{j_1} {{d^{j_1}}_{i_3}}^{j_2}{{d^{j_2}}_{i_4}}^{j_3} \ldots {{d^{j_{m-5}}}_{i_{m-3}}}^{j_{m-4}} {{d^{j_{m-4}}}_{i_{m-2} i_{m-1})}}\al_{i_1} \al_{i_2} \ldots \al_{i_{m-1}}&&
\cr
= \:  \frac{1}{2}\frac{1}{m!} {d_{(i_1i_2}}^{j_1} {{d^{j_1}}_{i_3}}^{j_2}{{d^{j_2}}_{i_4}}^{j_3} \ldots {{d^{j_{m-4}}}_{i_{m-2}}}^{j_{m-3}} {d_{i_{m-1}i_m)}}^{j_{m-3}} \al_{i_1}\al_{i_2} \ldots \al_{i_m}.&&
\label{eq:LHS-RHS-casimir-dijk-symm}
\eeqs
Thus, we see that pure states necessarily lie on the intersection of primitive Casimirs given by
\beq
    C^{(m)} = 2 m  \bigg(1-\frac{2}{N}\bigg) C^{(m-1)}.  
\label{eq:Pure-state-casimir}
\eeq
where
\beq
    C^{(3)} = g_{ijk} \al_{i}\al_{j}\al_{k} = 6 d_{ijk} \al_{i}\al_{j}\al_{k}, \quad C^{(2)} = g_{ij} \al_{i}\al_{j} = 2 \del_{ij} \al_{i}\al_{j}.
\label{eq:base-case-casimir}
\eeq
and $C^{(2)}$ satisfies (\ref{eq:pure-state-surface}) i.e. $C^{(2)} = 4(1-1/N)$.
Using above equation, we can explicitly characterize the Sudbery basis Casimir elements and their intersection contain the space of pure states $P_N$. For $m>N$, Casimirs are not primitive and they can be written in terms of lower order primitive Casimirs\cite{a-j-m-inv-ten}. 

We know for a space of $N \times N$ density matrices $D_N$, there are $N-1$ independent Casimirs. Since the total number of dimensions of the space is $N^2 - 1$, the intersection of all primitive Casimir surfaces is a symplectic submanifold of at most $N(N-1)$ dimensions. However, the dimensions of the symplectic submanifold do not always saturate to $N(N-1)$. For example, as we have shown in Section \ref{s:symplectic-pure-states}, pure states submanifold $P_N$ has only $2(N-1)$ real dimensions. In this section, we only showed that $P_N$ lies on the intersection of primitive Casimir surfaces given by (\ref{eq:Pure-state-casimir}). We didn't show the converse i.e. the intersection of the given primitive Casimir surfaces must only contain pure states. So, this might seem possible that the intersection of given primitive Casimirs is some higher dimensional symplectic submanifold, of which pure states only constitute a $2(N-1)$ dimensional subspace. However, as we show in the next Section \ref{s:den-mom}, the converse holds and intersection only contains pure states. Moreover, in the Section \ref{symp-leaves}, we show that the intersection has  exactly $2(N-1)$ dimensions.

Once we understand the geometrical picture of the full space of density matrices $D_N$, it is possible to extract symplectic 2-form $w$ on the space of pure states $P_N$ from the Poisson tensor $r$ of $D_N$  (\ref{eq:Mixed-Poisson-Tensor}). We don't show this explicitly but for a consistency check, we show  in the Appendix \ref{s:rec-geo} that the geometry induced on space $P_N$ from the geometry of $D_N$ is indeed same as the one we expect to be induced from $\mathbb{C}^{N}$ to $\mathbb{C}P^{N-1}$.

\subsection{Density matrix moments as Casimir surfaces}
\label{s:den-mom}
Here, we establish Tr$\rho^N$ as a Casimir. Using (\ref{eq:dens-defn}), we can write
\beq
\text{Tr}(\rho^m) = \sum_{r} \frac{1}{N^{m-r}}\binom{m}{r} \text{Tr}\left( \left(\sum \al_{k} T_{k}\right)^r \right)
= \sum_{r} \frac{1}{N^{m-r}}\binom{m}{r} \sum_{\{k_i\} = 1}^{N^2-1}\al_{k_1}\al_{k_2} \ldots \al_{k_r} \text{Tr}(T_{k_1}T_{k_2} \ldots T_{k_r}).
\eeq
Since $\al_i$s multiplies commutatively, we can write the above equation as symmetrized sums
\beq
\text{Tr}(\rho^m) = \sum_{r} \frac{1}{N^{m-r}}\binom{m}{r} \sum_{\{k_i\} = 1}^{N^2-1}\frac{1}{r!}\al_{k_1}\al_{k_2} \ldots \al_{k_r} \text{Tr}(T_{(k_1}T_{k_2} \ldots T_{k_r)}).
\label{eq:den-pol-sym-tr}
\eeq
where $T_{(k_1}T_{k_2} \ldots T_{k_r)} = \sum_j \prod_{i} T_{\sig_j{(k_i)}}$. This is an ad-invariant symmetric tensor satisfying (\ref{eq:G-invariance-identity}) (see Ref.~\onlinecite{a-j-m-inv-ten}). This implies $\al_{k_1}\al_{k_2} \ldots \al_{k_r} \text{Tr}(T_{(k_1}T_{k_2} \ldots T_{k_r)})$ is a Casimir. Consequently, Tr$\rho^m$ (\ref{eq:den-pol-sym-tr}) is also a Casimir. This provides us with another set of $N-1$ primitive Casimirs Tr$\rho^2$, Tr$\rho^3$, $\cdots$, Tr$\rho^N$. In fact, using (\ref{eq:prodTT}) and symmetrizing sums repeatedly, we can write the density matrix moments as multivariate polynomials in Sudbery basis Casimirs.

An orbit under the conjugate action by $U(N)$, i.e., $\text{orb}_{U(N)}(\rho)\: =\: \{U^{\dag} \rho U | U \in U(N)\}$ classifies all density matrices lying on a Casimir surface with constant Tr$\rho^m $. Since, a symplectic submanifold is precisely the intersection of such primitive Casimirs. Consequently, a symplectic submanifold in $D_N$ can be defined as an orbit of a diagonal matrix under the conjugate action by $U(N)$, since it preserves Tr$\rho^m$ for any $m$. They are also known as the coadjoint orbits.

Now, it is easy to see that pure states necessarily lie on the intersection of Casimir surfaces Tr$\rho^k = 1$. Conversely, any density matrix lying on the intersection of all primitive Casimirs Tr$\rho^k = 1$, must be a pure state. In fact, given the positive semi-definiteness, it is  sufficient to show that any density matrix satisfying Tr$\rho^2 = 1$ (and Tr$\rho = 1$) must be a pure state. In other words, the surface Tr$\rho^2 = 1$ inside $D_N^+$ must be $P_N$. This is easy to see as
\beq
\text{Tr} \rho = \sum_i d_i = 1 \quad \text{and} \quad \text{Tr} \rho^2 = \sum_i d_i^2 = 1
\label{eq:max-ent-pure}
\eeq
where $d_i$s are the eigenvalues of $\rho$.
Evaluating $\text{Tr} \rho -\text{Tr} \rho^2$, we get
\beq
\sum_i d_i (1 -  d_i) = 0.
\label{eq:diag-subt-pure}
\eeq

Since $d_i \geq 0$, each of the term in (\ref{eq:diag-subt-pure}) vanishes individually. This can only occur if $d_i=1$ or $0$. Additionally, $\sum_i d_i = 1$ guarantees that only one of the eigenvalues is $1$ while others are $0$, implying the density matrix must be a pure state. Thus, the Casimir Tr$\rho^2 = 1$ itself is symplectic manifold $P_N \subset D_N^+$ containing all pure states. It must also lie on the other Casimirs Tr$\rho^k = 1$ for $k > 2$. 

If we don't assume positive semi-definiteness, and consider the intersection of all primitive Casimirs of the form Tr$\rho^k = 1$ in $D_N$, their intersection is still $P_N$. This follows from a theorem discussed in the next Section \ref{s:Pos-cons-all-Cas-sur}, which guarantees that non-trivial intersection of a given $N-1$ primitive Casimir surfaces must be a sympletcic submanifold that either lie entirely inside $D_N^+$ or entirely outside $D_N^+$.  Since Tr$\rho^k = 1$ contains $P_N$ which lies inside $D_N^+$. Thus, the full intersection must be entirely inside $D_N^+$ implying the positive semi-definiteness. Hence, (\ref{eq:max-ent-pure}) and (\ref{eq:diag-subt-pure}) along with positive semi-definiteness proves the converse i.e the intersection of primitive Casimirs Tr$\rho^k = 1$ must contain only pure states. This establishes $P_N$ as a symplectic submanifold containing all pure states given by the intersection of primitive Casimirs Tr$\rho^k = 1$. 

\subsection{Positive semi-definiteness constraining the allowed Casimir surfaces.}
\label{s:Pos-cons-all-Cas-sur}

For any $N \times N$ Hermitian matrix $\rho$ to represent a positive semi-definite operator, it is necessary and sufficient for $S_m \geq 0$ for all $m$ where $S_m$ are coefficients of the characteristic equation for the Hermitian matrix $\rho$ (see Section  {\bf III A} of  Ref.~\onlinecite{Byrd-Khaneja}), i.e.
\beq
\text{det}(\rho - \la I) = \la^N - S_1\la^{N-1} + S_2 \la^{N-2} - \ldots + (-1)^N S_N = 0, 
\label{eq:charc-eq-dens}
\eeq
where
\beq
S_m = \frac{1}{m}\sum_{k=1}^{m} (-1)^{k-1} \text{Tr}(\rho^k)S_{m-k} \quad \text{and} \quad S_1 = \text{Tr}\rho = 1, \: S_0 = 1.
\label{eq:rec-pos-cas}
\eeq 
Since $S_m$ itself is a polynomial in Tr$\rho^k$, $S_m$ qualifies to be a Casimir. Thus, one may choose $\{S_2,S_3, \cdots, S_N\}$ as a set of primitive Casimir surfaces. The above theorem puts restrictions on the allowed values the Casimirs can take for the Hermitian matrices of trace $1$ to represent physically valid density matrices (positive semi-definite). For $m>N$, $S_m = 0$. This can be used to express non-primitive higher order Casimirs Tr$\rho^m$ for $m>N$ in terms of primitive Casimirs Tr$\rho^m$ for $ m \in \{2,3 \ldots N\}$. Note the theorem is valid for Hermitian matrices. Not all choices of $S_m \geq 0$ leads to
Hermitian matrices. For instance, choosing $S_2 = (1/2)(1 -\text{Tr}\rho^2) \geq 1/2 $ translates to $0 \geq \text{Tr}\rho^2$. Such a Casimir does not exist in real $D_N$ and the density matrix ceases to be Hermitian. Since for any positive semi-definite $\rho$, Tr$\rho^k \leq 1$ and $S_k$ is a multivariate polynomial in Tr$\rho^k$, $S_k$ for any physical density matrix is necessarily bounded.

Only a specific range of choices of primitive Casimirs $S_k \geq 0$ leads to a non-zero intersection in $D_N$. Using the above theorem, such an intersection must be a symplectic manifold containing only physical density matrices (i.e, positive semi-definite Hermitian matrices) and must necessarily lie in $D_N^+$. Similarly, if any $S_m < 0$, such a Casimir and all its symplectic submanifolds necessarily lie outside $D_N^+$. 
$\{ S_n, H\} = 0$ guarantees that $S_n$ remains unchanged under the time evolution. Thus, the trajectories in $D_N^+$ remain in $D_N^+$ and trajectories outside $D_N^+$ in $D_N$ remains outside $D_N^+$, for any arbitrary  Hamiltonian function $H$ defined on $D_N$. So, space of physical density matrices $D_N^+$ is a valid phase space.

Coming back to the density matrices $\rho$ satisfying Tr$\rho^k = 1$ imply they must satisfy $S_k = 0$ for $k \geq 2$ using (\ref{eq:rec-pos-cas}). Thus, by the above theorem, positive semi-definiteness is guaranteed for density matrices lying on the intersection of Tr$\rho^k = 1$. Thus, the arguments in previous Section \ref{s:den-mom} hold correct. And it establishes that the space of pure states $P_N$ is indeed a symplectic submanifold which is an intersection of Casimirs $S_k = 0$, or equivalently Tr$\rho^k = 1$ for $k \geq 2$.  

One may use any set of primitive Casimirs as per the convenience: $\{C^{(2)},\ldots,C^{(N)}\}$ (the Sudbery basis Casimirs), $\{\text{Tr}\rho^2,\text{Tr}\rho^3,\ldots, \text{Tr}\rho^N\}$ (density matrix moments) and $\{S_2,\ldots,S_N\}$ (coefficients of the characteristic polynomial). The Von-Neumann Entropy $S(\rho)=-\text{Tr}\rho\text{ln}\rho$ can be another useful Casimir to classify density matrices.

\subsection{Symplectic leaves of $D_N^+$} 
\label{symp-leaves}

Every density matrix $\rho$ lies in $\text{orb}_{U(N)} (y)$ where $y$ is some positive semi-definite diagonal matrix of trace $1$. The orbit of each such $y$ is precisely a symplectic submanifold (coadjoint orbits). We calculate the dimensions of such a symplectic submanifold in $D_N^+$. A general density matrix with trace 1 is given by
\beq
y(\{e_i,n_i\}) =  \begin{pmatrix}
 e_1 I_{n_1\times n_1} &  & &  &   \\
 & e_2 I_{n_2\times n_2}  &  &  &  \\
 &  & . &  &  \\
& &  & . &  \\
 &  &  & & e_k I_{n_k\times n_k} 
\end{pmatrix}_{N \times N}.
\label{eq:gen-daig-tr1}
\eeq
where $\sum_i d_i = \sum_i e_i n_i = 1$, $e_i \geq 0$ and $\sum_i n_i = N$. Using
\beq
\text{dim} \: \text{orb}_{U(N)}(x) = \text{dim} \: \left( U(N)/ U_x(N)\right) = \text{dim} (U(N)) - \text{dim}( U_x(N)),
\label{eq:orb-stab}
\eeq
  it is sufficient to calculate the dimensions of stabilizer of  (\ref{eq:gen-daig-tr1}). Stabilizer of (\ref{eq:gen-daig-tr1}) is given by $U(n_1) \times U(n_2) \times \cdots \times U(n_k)$ such that Unitary subgroups $I \times I \times \cdots U(n_i) \times I \cdots \times I$ act on different diagonal blocks. Thus, the number of dimensions of the stabilizer is given by $\sum_i n_i^2$. Now, we can calculate dimensions of a symplectic submanifold which is an orbit to some diagonal matrix given by (\ref{eq:gen-daig-tr1}):
\beq
\text{dim}\:\: \text{orb}_{U(N)}(y)= N^2 - \sum_i n_i^2 = (\sum_i n_i)^2 - \sum_i n_i^2 = 2 \sum_{i > j} n_i n_j. 
\label{eq:dim-symp}
\eeq
Thus, any orbit is even dimensional. The calculation of dimensions in (\ref{eq:dim-symp}) does not assume any constraints on the diagonal elements $e_i$ but only on the distribution $n_i$. So, the results hold generally for any symplectic submanifold in $D_N$. Constraints on $e_i$ are needed to classify the symplectic submanifolds of $D_N^+$. Therefore, we can write $D_N^+$ as a disjoint union of such symplectic submanifolds subject to the constraints:
\beq
D_N^+ = \bigsqcup_{\{n_i,e_i\}'} \text{orb}_{U(N)}(y\{n_i,e_i\})
\label{eq:disj-un}
\eeq
where $y\{n_i,e_i\}$ is defined by (\ref{eq:gen-daig-tr1}) and is constrained by $\sum_i e_i n_i = 1$, $e_i \geq 0$ and $\sum_i n_i = N$. Number of dimensions for each orbit of $y\{n_i, e_i\}$ is given by $2 \sum_{i>j} n_i n_j$. Since the dimensions of symplectic leaves can vary, the $D_N^+$ is said to be singularly foliated by the coadjoint orbits. The density matrix moments for coadjoint orbits to $y\{n_i,e_i\}$ are given by Tr$\rho^k = \sum_i n_i e_i^k$.
    
 For space of pure states $P_N$, the number of dimensions is simply $2(N-1)$ ($\{e_1 = 1, e_2 = 0\}, \{n_1 = 1, n_2 = N-1\}$). If all the diagonal elements are distinct (each $n_i = 1$), the number of dimensions is simply $2 \binom{N}{2} = N(N-1)$. Thus, when all diagonal elements are distinct, we recover the maximum dimensions of the symplectic submanifold that is possible. Each intersection by a primitive Casimir surface for such a case reduces exactly a single dimension. 
   
\section{Geometry of Entangled states}
\label{s:trace-geo}

\subsection{Tracing Geometry}
In this section, we trace out the states of $\mathbb{C}^M$ to get the geometry of $D_N^+$ from the space of pure states $\mathbb{C}^N \times \mathbb{C}^M$. Consider a generic pure state in $\mathbb{C}^N \times \mathbb{C}^M$:
\beq
\sum_{i,j} a_{ij} \:| i \ket_1 \otimes | j \ket_2
\label{eq:gen-entg-state}
\eeq  
such that $\{| i \ket_1\}$ and $\{| j \ket_2\}$ constitutes an orthonormal basis in $\mathbb{C}^N$ and $\mathbb{C}^M$ respectively. Following the results from \S\ref{s:symplectic-pure-states}, $a_{ij}$s furnish canonical coordinates for $\mathbb{R}^{2NM}$ i.e. $\{a_{ij}^r, a_{kl}^i\} = \del_{ik}\del_{jl}$ whereas $\{a_{ij}^r, a_{kl}^r\} = \{a_{ij}^i, a_{kl}^i\}=0$. The $r$ and $i$ denote real and imaginary part of $a_{ij}$ respectively. Writing (\ref{eq:gen-entg-state}) as a pure state density matrix:
\beq
\rho_{\text{pure}}= \sum_{i,j,k,l} a_{ij}a^*_{kl} \:| i \ket_1 \otimes | j \ket_2 \otimes  \bra l |_2 \otimes  \bra k|_1.
\label{eq:pure-dense-mat}
\eeq  
Tracing out the states in $\mathbb{C}^M$, we get a mixed state represented by $N \times N$ density matrix:
\beq
\rho= \sum_{i,j,k,l} a_{ij}a^*_{kl} \:| i \ket_1 \otimes \underbrace{\bra p | j \ket_2}_{\del_{pj}} \otimes  \underbrace {\bra l |p \ket_2}_{\del_{lp}} \otimes  \bra k|_1 = \sum_{i, k, p} a_{ip} a^*_{kp} \:| i \ket \otimes  \bra k|
\label{eq:traced-dense-mat}
\eeq  
where $\text{Tr} (\rho) = \sum_{ip} |a_{ip}|^2 $ and $\rho$ is Hermitian $( \:(\sum_{i, k, p} a_{ip} a^*_{kp})^* = \sum_{i, k, p} a_{kp} a^*_{ip})$. Using the Gell-Mann coordinates in the same way as in (\ref{eq:mix-stat-var-for-pure-gendim}) of Appendix \ref{s:rec-geo} to describe $N \times N$  density matrices, we get
\beq
\al_{s \neq 0} =  2\: \text{Tr}\: (\: \sum_{i, k, p} a_{ip} a^*_{kp} \:| i \ket \otimes  \bra k| \: T_s \:) \quad \text{and} \quad \al_0 = \text{Tr} \rho = \sum_{i, p} |a_{ip}|^2.
\label{eq:mixed-in-terms-of-entangled}
\eeq  
 The domain of the map (\ref{eq:mixed-in-terms-of-entangled}) is a real $2NM$ dimensional space of pure states and the co-domain is a $N^2$ real dimensional space of density matrices. Since (\ref{eq:mixed-in-terms-of-entangled}) is of same form as (\ref{eq:dense-mat-for-pure}) up to the sum over $p$, we recover the exactly same Poisson manifold $(A_N, r^{ij})$ after the pushforward as in Appendix \ref{s:rec-geo}:
\beq
 \{\al_s,\al_t\} = 4 f_{stu} \al_u \quad \text{for} \quad s,t \neq 0 \quad \text{and} \quad  \{\al_0,\al_t\} = 0 \quad \forall t
\label{eq:poiss-ent-tra}
\eeq
For representing a normalized state, we restrict to Casimir $\al_0 = \text{Tr}\rho =  \sum_{i, j} |a_{ij}|^2 = 1$, which is precisely $D_N^+$. The density matrices here are constructed as a partial trace over states in $\mathbb{C}^M$ (\ref{eq:mixed-in-terms-of-entangled}) while the full system lives in $\mathbb{C}^N \times \mathbb{C}^M$. The trace of $\rho^m$ for this density matrix is given by
\beq
\text{Tr}(\rho^m) = a_{i_1 j_1} a^*_{i_2 j_1}a_{i_2 j_2} a^*_{i_3 j_2}a_{i_3 j_3} a^*_{i_4 j_3} \cdots a_{i_{m-1} j_{m-1}} a^*_{i_m j_{m-1}}a_{i_m j_m} a^*_{i_1 j_m}
\label{eq:trarhom}
\eeq
The above expression is independent of the order we trace, i.e.
\beq
 \text{Tr}_1(\text{Tr}_2\sum_{i,j,k,l} a_{ij}a^*_{kl} \:| i \ket_1 \otimes | j \ket_2 \otimes  \bra l |_2 \otimes  \bra k|_1)^m = \text{Tr}_2(\text{Tr}_1\sum_{i,j,k,l} a_{ij}a^*_{kl} \:| i \ket_1 \otimes | j \ket_2 \otimes  \bra l |_2 \otimes  \bra k|_1)^m.
\label{eq:tr-inv-dense-mat}
\eeq  
Let's assume $M<N$ and we trace out the states in $\mathbb{C}^N$ to get a collection of $M \times M$ density matrices $\rho$. As discussed in Section \ref{s:casimir-invariants}, it is sufficient to specify primitive Casimirs Tr$\rho^2$, Tr$\rho^3 \cdots ,$Tr$\rho^M$ to uniquely specify a symplectic submanifold. On the other hand, if we trace out the states in $\mathbb{C}^M$, we get a collection of $N \times N$ density matrices $\rho$. However, we no longer need to specify all $N-1$ primitive Casimirs Tr$\rho^2$, Tr$\rho^3$,$\cdots$, Tr$\rho^N$. Since from (\ref{eq:trarhom}) and (\ref{eq:tr-inv-dense-mat}), Tr$\rho^k$ is same for either order of tracing. In both spaces, it is sufficient to specify only Tr$\rho^2, \cdots,$ Tr$\rho^M$ and we can determine all other Casimirs Tr$\rho^k$ using $S_k = 0$ for $k>M$ from (\ref{eq:rec-pos-cas}). In other words, any $N \times N$ density matrix constructed by tracing over states of $\mathbb{C}^M$ necessarily lies on the intersection of $N-M$ primitive Casimir surfaces given by $\{ S_{M+1} = 0, S_{M+2} = 0, \cdots, S_{N} = 0 \}$. The converse may also be shown to hold in $D_N^+$ by calculating the rank of such density matrices and using the purification theorem. 

We will refer to these intersections of $N-M$ Casimirs in $D_N^+$ as $D_N^M$. The characteristic equation of such density matrices is given by  
\beq
\text{det}(\rho - \la I) = \la^{N-M}(\la^M - S_1\la^{M-1} + S_2 \la^{M-2} -  ... + (-1)^M S_M).
\label{eq:charc-eq-dens-CNxCM}
\eeq 
Thus, $N-M$ is the nullity of density matrices of $D_N^M$. The dimensions of $D_N^M$ cannot be more than $N(N-1)+(M-1)$. For $M>1$, the induced Poisson geometry on $D_N^M$ is not symplectic and still equipped with an induced degenerate Poisson tensor. In fact, $D_N^1 \subset D_N^2  \subset \cdots \subset D_N^{N-1}\subset D_N^N$, where $D_N^1 = \mathbb{C}P^{N-1}$ is symplectic and $D_N^N = D_N^+$ is a full space of physical $N \times N$ density matrices. We may calculate exact dimensions of physical density matrices $D_N^M$ defined by (\ref{eq:traced-dense-mat}) which maybe rewritten as 
\beq
\rho = \sum_{i, k = 1}^N  \bra \vec{a}_k| \vec{a}_i \ket \:| i \ket \otimes \bra k|
\quad \text{such that} \quad \text{Tr} \rho = \sum_i|\bra \vec{a}_i| \vec{a}_i \ket| = 1
\label{eq:mixed-vec-not-n>2}
\eeq  
where $\vec{a}_j = (a_{j_1}, a_{j_2}, \cdots ,a_{j_M})$ is an $M$-dimensional complex vector and $\bra \vec{a}_k| \vec{a}_i \ket$ is the complex inner product. This shows physical density matrices in $D_N^M$ can completely be determined in terms of inner products $\bra \vec{a}_k| \vec{a}_i \ket$. Since $\vec{a}_j$ lives in an $M$-dimensional complex vector space, we can choose basis $B = (\vec{a}_1,\vec{a}_2,\cdots, \vec{a}_M)$.  If we know the inner products between all the basis elements and the inner products of $\{\vec{a}_{M+1},\vec{a}_{M+2},\cdots,\vec{a}_N\}$ with the basis elements, it is sufficient to determine any inner products between the given vectors. Combined with the fact that Tr$\rho = 1$, 
\beq
\text{dim}( D_N^M) = \underbrace{2\binom{M}{2} + \binom{M}{1}  - 1}_{M^2-1} + 2(N-M)M  = 2NM-M^2 -1\quad \text{for} \quad N \geq M \geq 1.
\label{eq:dim-DNM}
\eeq  
At $M=1$, we recover the dimensions of $P_N$ as $2(N-1)$. At $M=N$, we recover the dimensions of $D_N^+$ as $N^2-1$. We note $N(N-1)+(M-1) \geq 2NM-M^2 -1$ (from $(N-M)^2 \geq (N-M)$ upon rearranging). Thus, the expected dimensions are not saturated for $M < N$. In the same way as (\ref{eq:disj-un}), $D_N^M$ is also foliated by the symplectic leaves:
\beq
D_N^M = \bigsqcup_{\{n_i,e_i\}'} \text{orb}_{U(N)}(y\{n_i,e_i\})
\label{eq:disj-un-sub-pois}
\eeq
 where $\{e_1 = 0, n_1 = N-M\}$. The sum is constrained by $\sum_{i > 1} n_i = M$ and  $\sum_{i > 1} n_i e_i =1 $. Note that $D_N^M$ can be assigned $M^2-1$ coordinates which form $\mathfrak{su}(M)$ Lie algebra. But it also has additional $2M(N-M)$ degrees of freedom. 
 
 For $M \geq N$, $D_N^M = D_N^+$ i.e. $D_N^N = D_N^{N+1} = D_N^{N+2} = \ldots = D_N^+$. This follows from following argument. $D_N^+$ is a collection of all $N \times N$ density matrices. So, for any $M$, $D_N^M \subseteq D_N^+$. It turns out that converse also holds for $M \geq N$. It follows from the purification theorem. Any $N \times N$ density matrix can be written as $\rho = \sum_{i=1}^{N} d_i | i\ket_1 \otimes\bra i|_1$ such that $d_i \geq 0$ and $\sum_{i=1}^N d_i = 1$. The $\{ | i\ket_1 \}$ constitutes some normalized basis for $\mathbb{C}^N$ which is not necessarily orthogonal. Such a density matrix can be obtained by partial tracing of any entangled state of the form $\sum_{i=1}^{N} \sqrt{d_i} | i \ket_1 \otimes | j(i) \ket_2$ where $\{ | j(i) \ket_2 \}$ constitutes an orthonormal basis. Thus, for this to be possible, $\{ | j(i) \ket_2 \}$ has to span at least $N$ dimensional Hilbert-space. The happens for $M \geq N$. Hence, any given $N \times N$ density matrix can be constructed by partially tracing over some state in $\mathbb{C}^N \times \mathbb{C}^M$ for $M \geq N$. This implies  $ D_N^+ \subseteq D_N^M$. So, $D_N^+ = D_N^M$ for all $M \geq N$.

\subsection{Maximally Entangled states in $D_N^M$}

Classifying all primitive Casimirs is sufficient to specify a unique symplectic submanifold. Specifying physically valid $S_k \geq 0$ for $k \in {2,3,..,N}$ carves us a symplectic manifold consisting of physical density matrices with same eigenvalues. However, as we saw in case of $P_N$, specifying all primitive Casimirs is not always necessary.  Specifying Tr$\rho^2 = 1$ along with positive semi-definiteness was sufficient to pick out physical Hilbert space. All other primitive Casimirs were automatically determined. Here, we look into a class of symplectic submanifolds consisting of density matrices having maximum entropy Ln$M$ in $D_N^M$. Let's call this $E_N^M$. The eigenvalues of such density matrices are either $1/M$ or $0$. We can calculate the primitive Casimirs on whose intersection the $E_N^M$ lies. They are given by $\text{Tr}\rho^k = 1/M^{k-1}$. But even if we just provide the 2 primitive Casimirs Tr$\rho^2 = 1/M$ and Tr$\rho^3 = 1/M^2$ and demanded the positive semi-definiteness, the intersection of these two Casimirs exactly carve out the space $E_N^M$. To show this, let $d_i \geq 0$ be the eigenvalues of $\rho$. Then
\beq
\text{Tr} \rho = \sum_i d_i = 1, \quad \text{Tr} \rho^2 = \sum_i d_i^2 = \frac{1}{M} \quad \text{and} \quad\text{Tr} \rho^3 = \sum_i d_i^3 = \frac{1}{M^2}.
\label{eq:max-ent}
\eeq
Evaluating $M \text{Tr} \rho - M^2 \text{Tr} \rho^2$ and  $M^2 \text{Tr} \rho^2 - M^3 \text{Tr} \rho^3$ in (\ref{eq:max-ent}), we get
\beq
\sum_i M d_i (1 - M d_i) = 0 \quad \text{and} \quad \sum_i M^2 d_i^2 (1 - M d_i) = 0.
\label{eq:diag-subt}
\eeq
Again subtracting the second equation from the first, we get
\beq
\sum_i M d_i (1 - M d_i)^2 = 0.
\label{eq:subt}
\eeq
Since $d_i \geq 0$, each of the term vanishes individually. This can only happen if $d_i = 1/M$ or $d_i = 0$. Thus, the intersection of 2 Casimirs must be the symplectic manifold $E_N^M \subset D_N^M$ i.e $E_N^M =$\{Tr$\rho^2 = 1/M$\} $\cap$ \{Tr$ \rho^3 = 1/M^2$\} $\cap$ $D_N^+$. The dimensions can be calculated using (\ref{eq:dim-symp}) which is given by $2M(N-M)$. The manifold lies on the intersection of Casimirs Tr$\rho^k = 1/M^{k-1}$. Geometrically, this means that the only primitive Casimirs of the form Tr$\rho^k$ for $k > 3$ to have any intersection with \{Tr$\rho^2 = 1/M$\} $\cap$ \{Tr$\rho^3 = 1/M^2$\} in $D_N^+$ are Tr$\rho^k = 1/M^{k-1}$. Any other choice for Tr$\rho^k$ for $k>3$ other than $1/M^{k-1}$ is an empty intersection in $D_N^+$. On the other hand, the intersection of  Tr$\rho^k = 1/M^{k-1}$ for $k > 3$ with \{Tr$\rho^2 = 1/M$\} $\cap$ \{ Tr$\rho^3 = 1/M^2$\} in $D_N^+$ is such that it fully contains the latter, without carving out any extra space.

\section{Discussion}

In this paper, we have developed the Poisson geometric formulation of quantum mechanics for finite dimensional density matrices. The formulation naturally extends the quantum phase space geometry to include both pure and mixed states on the same footing. Time evolution is generated by the Hamiltonian function Tr$(\rho \hat{H})/2 \hbar$. The Poisson manifold admits a linear Poisson (Kirillov-Kostant-Souriau) structure with corresponding lie algebra $\mathfrak{su}(N)$ and is shown to get foliated by symplectic leaves (coadjoint orbits). The space of pure states $\mathbb{C}P^{N-1}$ is one of its symplectic submanifolds. Different kinds of Casimirs are identified whose intersections can be used to specify its symplectic submanifolds. The Poisson structure also appears via partial tracing of the states in $\mathbb{C}^N \times \mathbb{C}^M$ induced from its symplectic structure. By varying $M$, we get a chain of Poisson submanifolds related by inclusion, each lying on intersection of some primitive Casimirs. Larger $M$ allows symplectic leaves consisting of mixed states with larger entropy and lie on the intersection of smaller number of primitive Casimirs. Their foliations are discussed as $M$ is varied. At $M=1$, we get only pure states and at any $M \geq N$, we get all possible $N \times N$ density matrices. Thus, it is possible to see how various degrees of entanglement manifest themselves as various Poisson geometric sub-structures in the space of density matrices. 

The density matrix phase space can also be thought of as a generalized classical mechanical angular momentum phase space. The tracing map (\ref{eq:mixed-in-terms-of-entangled}) is analogous to the definition of classical angular momenta in terms of positions and momenta. The density matrix phase space inherits $\mathfrak{su}(N)$ linear Poisson structure from the K\"ahler structure of $\mathbb{C}^N \times \mathbb{C}^M$ via partial tracing for $M \geq N$. This is quite analogous to how angular momentum phase space inherits $\mathfrak{su}(2)$ Poisson structure from the canonical sympletcic structure of classical phase space of positions and momenta.

For $M < N$, $D_N^M$ has more interesting Poisson structure. The dimensional calculation (\ref{eq:dim-DNM}) suggests the coordinates for $D_N^M$ may be broken into two sets with $M^2-1$ and $2M(N-M)$ number of coordinates. The first set may form $\mathfrak{su}(M)$ lie algebra and the second set may be assigned canonical coordinates, as it looks like from the symplectic structure of $E_N^M$. It will be an insightful exercise to assign coordinates to $D_N^M$ or its symplectic leaves and study their Poisson commutation relations.

Positive semi-definiteness has been shown to play a special role in the intersection of Casimirs. It is shown that physical density matrices (positive semi-definite Hermitian matrices of unit trace) constitute a valid phase space for any arbitrary Hamiltonian function. Moreover, positive semi-definiteness often reduces the requirement to specify all the primitive Casimirs to specify a symplectic submanifold. This has been shown for the pure states $P_N$ and the mixed states with largest entropy $E_N^M$ in each $D_N^M$. It will be interesting to learn the positive semi-definite connection with the intersection of Casimir surfaces in more generality.

 The formalism developed in this paper is restricted to finite dimensions. If we could extend the formalism to infinite dimensions, this will allow us to include mixed states in the geometric picture developed in Refs.~\onlinecite{AbAs-Tras,Bojowald-Skirzewski-eff, Bojowald-Baytas-Crowe}. Upon inclusion, the quantum fibre-bundle phase space will contain a far wider class of quantum mechanical states. It will be very interesting to see how the overall geometry unfolds once we include the mixed states and what are its implications for semi-classical physics. This direction of study is important as it may help us address some thorny issues like defining the chaos itself in quantum mechanics. The notion of chaos in classical mechanics is defined by sensitive dependence to initial conditions. The exponential divergence of nearby trajectories provides us the largest Lyapunov exponent, a well known measure of classical chaos. Although, there are different measures of chaos in quantum mechanics, e.g. Loschmidt echo, out of time ordered correlators, where we find a connection with the Lyapunov exponents of a classically chaotic system, a more direct relation between the chaos in both regimes is yet to be understood. In the geometric formalism, since we can define trajectories for a quantum system, we can extend the classical mechanical definition of chaos. Recently, the effective methods have been used to study the chaos for the semi-classical Mixmaster models \cite{martin-sara, martin-sara-2}. Examination of such extensions may provide us with a more general and may be a unified view of manifestation of chaos. 

In this paper, we have restricted our attention to the Poisson geometric structure only, which has a classical analog. We do not touch upon the Riemannian geometric structure, which has been emphasized to be equally important and is responsible for the aspects of quantum mechanics that are absent in classical mechanics. So, if we really want to address issues in the semi-classical regime, it will not be wise to leave out the Riemannian geometric picture of quantum mechanics. Thus, one important direction is to explore the possibility of how the full K\"ahler geometry of pure states carries over to the space of density matrices and what does it signify? The ultimate aim of course will be to extend all of them to infinite-dimensions. This will allow us to address semi-classical physics and quantum-mechanical generalizations with full geometric machinery. 

\section*{Acknowledgements}

We thank Alok Laddha for suggesting this direction of work and for helpful discussions and comments. We thank Martin Bojowald for helpful comments and for suggesting useful references.
 
\appendix

\section{Proof of ad-invariance identity}
\label{s:ad-inv}

We will first start by proving the ad-invariance identity (\ref{eq:G-invariance-identity}) for $m=3$. In this case the ad-invariance identity simply follows from the second Jacobi identity:
\beq
[\{T_a, T_b\},T_c] + [\{T_b,T_c\},T_a] + [\{T_c,T_a\},T_b] = 0,
\eeq
where $\{T_i,T_j\} = T_i T_j + T_j T_i$.

For convenience, we will use the adjoint representation of $\mathfrak{su}(N)$ where, the generators of $\mathfrak{su}(N)$ are denoted by $
(N^2-1) \times (N^2-1)$ anti-symmetric matrices $F_a$. The matrix elements of $F^a$ are
\beq
(F_a)_{bc} = -i f_{abc}.
\eeq
We also define $(N^2-1) \times (N^2-1)$ traceless symmetric matrices
\beq
(D_a)_{bc} = d_{abc}.
\eeq
In terms of these, the second Jacobi identity becomes
\beq
[F_a,D_b] = -i f_{abc}D_c.
\eeq
Writing it in component form, we obtain the ad-invariance identity for $m=3$
\beq
 f_{mlj} d_{npj}  + f_{nlj} d_{pmj}  +  f_{plj} d_{mnj} = 0.
\eeq
Similarly, we can obtain the ad-invariance identity for any $m$ by considering
\beq
[F_a, D_{(b_1} D_{b_2} \cdots D_{b_{m-2})}]= -i \sum_{i=1}^{m-2} f_{ab_i c} D_{(b_1} D_{b_2} \cdots D_{\hat{b}_i c} \cdots D_{b_{m-2})}.
\eeq
Again, by writing it in component form and relabeling the indices we get the ad-invariance identity in the form of (\ref{eq:G-invariance-identity}).

\section{Reconciling the Geometries}
\label{s:rec-geo}

In this appendix, we show the geometry induced on space $P_N$ from the geometry of $D_N$ is indeed same as the one we expect to induce from $\mathbb{C}^{N}$ to $P_N$. The finite dimensional physical Hilbert space $P_N = \mathbb{C} P^{N-1}$ is constructed by identifying $(z_1, z_2,\cdots,z_N)$ with $\la (z_1, z_2,\cdots,z_N)$ where $\la$ can be any complex number. To go from $\mathbb{C}^N$ to $\mathbb{C} P^{N-1}$, one may map a pure state to a density matrix itself as an intermediate step: 
\beq
\begin{pmatrix}
 z_1\\
 z_2 \\
.\\
.\\
z_N
\end{pmatrix}
\:\:
\to
\:\:
\begin{pmatrix}
 z_1\\
 z_2  \\
. \\
. \\
z_N
\end{pmatrix}
\begin{pmatrix}
 z_1^* & z_2^* & . & . & z^*_N   
\end{pmatrix}
 =
\begin{pmatrix}
 |z_1|^2 &  z_1 z_2^* & . & . & z_1z^*_N\\
  z_2 z_1^* & |z_2|^2& . & . & z_2z^*_N \\
. & . & . & .  & .\\
. &. & . & . & . \\
  z_N z_1^* & z_N z_2^*& . & . & |z_N|^2 
\end{pmatrix}. 
\label{eq:pureCntomixed}
\eeq  

The pure state written in form of density matrix is invariant under the action of $U(1)$. So, arbitrariness of overall phase angle is taken care of. Moreover, we may use Gell-Mann coordinates themselves to represent these states, i.e.
\beq
 \rho = \frac{\al_0}{N}I + \sum_{k=1}^{N^2-1} \al_k T_k=
 \sum_{i,j=1}^{N} z_i z^*_j \:| i \ket \otimes  \bra j|.
\label{eq:mix-stat-var-for-pure-gendim}
\eeq  
Using above, we can define our new coordinates as
\beq
\al_{s \neq 0} =  2\: \text{Tr}\: (\: \sum_{i, j} z_i z^*_j \:| i \ket \otimes  \bra j| \: T_s \:) = 2  \sum_{i, j} z_i z^*_j \:\bra j| \: T_s | i \ket  \quad \text{and} \quad \al_0 = \text{Tr} \rho = \sum_i |z_i|^2.
\label{eq:dense-mat-for-pure}
\eeq  
Note that unlike (\ref{eq:dens-defn}), the definition (\ref{eq:mix-stat-var-for-pure-gendim}) treats $\al_0$ as a new variable in Gell-Mann coordinates.
If the coordinates are allowed to take arbitrary real values, they constitute a $N^2$ dimensional real manifold $A_N$ (space of auxiliary density matrices). Note the map (\ref{eq:dense-mat-for-pure}) from $\mathbb{C}^N \to A_N$ is neither one-to-one nor onto. Any element in the $\text{orb}_{U(1)}(\vec{z})$ is mapped to same density matrix. But since the physical states live in $\mathbb{C}P^{N-1}$, there exist a distinct density matrix for each physical state. Moreover, any physical state is still determined in $A_N$ up to an overall real scaling factor. The image of the $\mathbb{C}^N$ under this map is a $2N-1$ dimensional real manifold. From (\ref{eq:pureCntomixed}), the image can be parameterized by $2N-1$ variables $\{|z_j|, \tht_k - \tht_{k+1} | 1 \leq j \leq N, 1 \leq k \leq N-1 \}$ where $\tht_k = \text{Arg}(z_k)$. This is because any entry $z_i z^*_j$ in (\ref{eq:pureCntomixed}) can be written as $|z_i|  |z_j| \exp(\tht_i - \tht_j)$ where $\tht_i - \tht_j$ (for $i < j$) is just the sum $(\tht_i - \tht_{i+1})+(\tht_{i+1} - \tht_{i+2})+ \cdots +(\tht_{j-1} - \tht_j )$. The image lives as a subset in a $N^2$ real dimensional manifold $A_N$ parameterized by $\al_0, \al_1, \cdots, \al_{N^2-1}$. The Poisson tensor $r^{ij}$ on $\mathbb{C}^N$ can be pushed forward to this new space  where we find the Poisson algebra is closed in new variables. 
We show this by explicitly calculating the Poisson commutation relations for the new coordinates $\al$ defined by (\ref{eq:dense-mat-for-pure}). Since $T$ is Hermitian, 
writing $\al_s$ (for $s \neq 0$) in terms of canonical coordinates, we get
\beq
\al_s =  2  \sum_{i, j = 1}^{N} (x_i x_j + y_i y_j) T^{ji}_{sR} - (y_i x_j - x_i y_j) T^{ji}_{sI}
\label{eq:aux-in-term-of-can}
\eeq  
where
\beq
 T^{ji}_{sR} = \Re(\bra j|  T_s | i \ket) \quad \text{and} \quad  T^{ji}_{sI} = \Im(\bra j|  T_s | i \ket) .
\label{eq:re-and-im-gelman}
\eeq  
Since $T$ is Hermitian ($\bra j|  T_s | i \ket^* = \bra i |  T_s | j \ket$), we get the relations
\beq
 T^{ji}_{sR} =  T^{ij}_{sR}; \quad T^{ji}_{sI} =  -T^{ij}_{sI}.
\label{eq:T-hermitian}
\eeq  
We calculate $\{\al_s,\al_t\}$ to get
 \beq
 4 \{ (x_i x_j + y_i y_j) T^{ji}_{sR} - (y_i x_j - x_i y_j) T^{ji}_{sI}, (x_k x_l + y_k y_l) T^{lk}_{tR} - (y_k x_l - x_k y_l) T^{lk}_{tI}\}.
\label{eq:als-alt}
\eeq  
We use the commutation relations $(\{x_i, y_j\} = -\{y_j, x_i\}= \del_{ij}, \{x_i, x_j\} = \{y_i, y_j\} = 0)$ to calculate (\ref{eq:als-alt}). Using the linearity of Poisson brackets, we break the calculation into 4 parts. The first term is given by
  \beqs
 \{ (x_i x_j + y_i y_j) T^{ji}_{sR}, (x_k x_l + y_k y_l) T^{lk}_{tR}\} = \{ x_i x_j ,y_k y_l\}  (T^{ji}_{sR}T^{lk}_{tR} - T^{lk}_{sR}T^{ji}_{tR}).
\label{eq:als-alt-1st-term}
\eeqs  
Here, we relabelled the indices and  used the anti-commutativity. Now using the Leibniz rule and commutation rules, we get
 \beq
 \{ (x_i x_j + y_i y_j) T^{ji}_{sR}, (x_k x_l + y_k y_l) T^{lk}_{tR}\}  =  4 T^{ik}_{sR}T^{kj}_{tR} (x_i y_j - y_j x_i).
\label{eq:als-alt-1st-term-2}
\eeq  
The fourth term is given by 
 \beqs
  \{(y_i x_j - x_i y_j) T^{ji}_{sI},(y_k x_l - x_k y_l) T^{lk}_{tI}\} = 4 T^{ji}_{sI}T^{lk}_{tI} \{ y_i x_j ,y_k x_l\}.  
\label{eq:als-alt-4th-term}
\eeqs  
Here, we relabelled the indices and used the anti-symmetry of $\Im T$. Now, we use Leibniz rule and compute the commutation relations to get 
\beqs  
  \{(y_i x_j - x_i y_j) T^{ji}_{sI},(y_k x_l - x_k y_l) T^{lk}_{tI}\} = 4 T^{ji}_{sI}T^{lk}_{tI} (y_i x_j - x_i y_j).
\label{eq:als-alt-4th-term-2}
\eeqs  
Calculating the third term, we get 
\beq
- \{(y_i x_j - x_i y_j) T^{ji}_{sI}, (x_k x_l + y_k y_l) T^{lk}_{tR}\} = - 2 T^{ji}_{sI}T^{lk}_{tR} \{y_ix_j, x_k x_l + y_k y_l\}.
\label{eq:als-alt-3rd-term}
\eeq
Here, we again used the anti-symmetry of $\Im T$. After this, we use Leibniz rule and compute the commutation relation. We again relabel and use the anti-symmetry to get
\beq
- \{(y_i x_j - x_i y_j) T^{ji}_{sI}, (x_k x_l + y_k y_l) T^{lk}_{tR}\} = 4 T^{ik}_{sI}T^{kj}_{tR}(x_i x_j + y_i y_j) 
\label{eq:als-alt-3rd-term-2}
\eeq
For calculating 2nd term, we note that 
\beq
-2 \{ (x_i x_j + y_i y_j) T^{ji}_{sR}, (y_k x_l - x_k y_l) T^{lk}_{tI}\} \longleftrightarrow -(\ref{eq:als-alt-3rd-term}) \equiv s  \longleftrightarrow  t.
\
\label{eq:als-alt-2nd-term}
\eeq
So, second term is
\beq
- \{(y_i x_j - x_i y_j) T^{ji}_{sI}, (x_k x_l + y_k y_l) T^{lk}_{tR}\} =  4 T^{ik}_{sR}T^{kj}_{tI}(x_i x_j + y_i y_j). 
\label{eq:als-alt-2nd-term-2}
\eeq
Here, we again used the anti-symmetry to write the final form. 
Collecting (\ref{eq:als-alt-1st-term-2}), (\ref{eq:als-alt-4th-term-2}), (\ref{eq:als-alt-3rd-term-2}) and (\ref{eq:als-alt-2nd-term}), we get
\beq
\{\al_s,\al_t\} = 16 [(y_i x_j - x_i y_j)(T^{ik}_{sR}T^{kj}_{tR} - T^{ik}_{sI}T^{kj}_{tI})+ (x_i x_j + y_i y_j)(T^{ik}_{sR}T^{kj}_{tR} + T^{ik}_{sI}T^{kj}_{tI})].
\label{eq:als-alt-LHS}
\eeq
The RHS of above equation can be written as imaginary part of the product of 2 terms i.e.
\beq
\text{RHS} = 16 \Im \left[ \left((x_i x_j + y_i y_j)+ i(y_i x_j - x_i y_j)\right)
\left((T^{ik}_{sR}T^{kj}_{tR} - T^{ik}_{sI}T^{kj}_{tI})+i(T^{ik}_{sR}T^{kj}_{tR} + T^{ik}_{sI}T^{kj}_{tI})\right)  \right].
\label{eq:als-alt-RHS-1}
\eeq
This may further be broken into the products.
\beq
\text{RHS} = 16 \Im \left[ (x_i + i y_i)(x_j - i y_j)(T^{ik}_{sR}+iT^{ik}_{sI})(T^{ik}_{tR}+iT^{kj}_{tI})\right].
\label{eq:als-alt-RHS-2}
\eeq
Thus, we get
\beq
\text{RHS} = 16 \sum_{i,j,k }\Im \left[ z_i z^*_j \bra j |  T_s | k \ket \bra k | T_t | i \ket  \right] = 16 \sum_{i,j}\Im \left[ z_i z^*_j \bra j |  T_s T_t | i \ket  \right].
\label{eq:als-alt-RHS-3}
\eeq
Writing the imaginary part as
\beq
\text{RHS} = -8i \sum_{i,j}  [z_i z^*_j \bra j |  T_s T_t | i \ket  - z_j z^*_i \bra j |  T_s T_t | i \ket^*] = -8i \sum_{i,j}  z_i z^*_j \bra j |  [T_s ,T_t ] | i \ket.
\label{eq:als-alt-RHS-4}
\eeq
Here, we relabelled the indices and used the Hermiticity of $T_s$ and $T_t$ to get the above. Now, using $ [T_s ,T_t ] = i f_{stu} T_u$, we get 
\beq
\text{RHS} = -8 \sum_{i,j} f_{stu} z_i z^*_j \bra j |  T_u| i \ket = 4  f_{stu} \al_u.
\label{eq:als-alt-RHS-5}
\eeq
Thus, we recover $\mathfrak{su}(N)$ lie algebra $\{\al_s,\al_t\} = 4 f_{stu} \al_u$ for $s \neq 0$. For $s =0$, it is easy to show $\{\al_0, \al_s\}$ is identically $0$.
\beq
\{\al_0, \al_s\} =  2\sum_{i, j,k} \{x_k x_k + y_ky_k,  (x_i x_j + y_i y_j) T^{ji}_{sR} - (y_i x_j - x_i y_j) T^{ji}_{sI}\}.
\label{eq:a0centre}   
\eeq
Coefficient of $T^{ji}_{sR}$ and $T^{ji}_{sI}$ individually commutes with $\al_0$ for each $i,j$ i.e.
\beq 
\{x_k x_k + y_ky_k, x_i x_j + y_i y_j\} = 2 (x_iy_j + x_jy_i - y_ix_j - y_jx_i) = 0
\eeq
and
\beq 
\{x_k x_k + y_ky_k, y_i x_j - x_i y_j\} = 2 (x_ix_j - x_jx_i - y_iy_j + y_jy_i) = 0.
\eeq
This implies $\{\al_0, \al_s\}  = 0$ in (\ref{eq:a0centre}) for all $s$. Hence, $\al_0$ lies in the center of lie algebra $\mathfrak{u}(N)$ spanned by $\al_0, \al_1 .. \al_{N^2-1}$. Moreover, $\al_0 = c$ is the most trivial family of Casimir surfaces in $(A_N, r^{ij})$ where $c$ takes real values.

Since a pure state density matrix $\rho$ in (\ref{eq:pureCntomixed}) satisfies $\rho^2 = (\sum_i |z_i|^2) \rho$. So, for uniquely characterizing the $2(N-1)$ dimensional space of physical pure states, we get rid of the real scaling factor by imposing $\sum_i |z_i|^2 = 1$. In terms of Gell-Mann coordinates $\al$s describing $A_N$, from (\ref{eq:rho-square}) and (\ref{eq:mix-stat-var-for-pure-gendim}), the normalization implies $\al_0 = 1$ and $\sum_k\al_k \al_k = 2(1-1/N)$. This fixes our two Casimir surfaces. $\al_0 = 1$ brings us to a $N^2-1$ dimensional space exactly same as $(D_N, r^{ij})$ where $r^{ij}$ is the Poisson algebra of the Lie algebra $\mathfrak{su}(N)$ (\ref{eq:als-alt-RHS-5}). The surface $\sum_k\al_k \al_k = 2(1 - 1/N)$ at $\al_0 = 1$ is same as the Casimir surface (\ref{eq:pure-state-surface}). Moreover, upon normalizing, the $\rho^2 = \rho$ is automatically satisfied by (\ref{eq:pureCntomixed}) implying the same set of constraints on Gell-Mann variables as in Section \ref{Pure-stat-casimir}. Thus, the space of physical pure states $(\mathbb{C}P^{N-1})$ with the induced geometry from $\mathbb{C}^N$ is same as the pure states space $(P_N)$ of Section \ref{Pure-stat-casimir} with the induced geometry from $D_N$. 


\end{document}